\documentstyle[12pt,psfig]{article}
\def\be{\begin{equation}}
\def\ee{\end{equation}}                                                                                                         

\begin{document}
\begin{titlepage}

\begin{center} {\large \bf Reaction Diffusion Models Describing a Two-lane 
Traffic Flow  
 } \\

\vskip 1cm

\centerline {\bf M. Ebrahim Fouladvand \footnote
{e-mail:foolad@theory.ipm.ac.ir} } \vskip 1cm
{\it  Department of Physics, Sharif University of Technology, }\\
{\it P.O.Box 11365-9161, Tehran, Iran }\\
and\\
{\it  Institute for Studies in Theoretical Physics and Mathematics,}\\
{\it P.O.Box 19395-5531, Tehran, Iran}
\end{center}

\begin{abstract}

A uni-directional two-lane road is approximated by a set of two parallel 
closed one-dimensional chains. Two types of cars i.e. slow and fast ones
are 
considered in the system. Based on the {\it Nagel-Schreckenberg } (Na-Sch)
model of traffic flow \cite{NS92}, a set of reaction-diffusion processes
are introduced 
to simulate the behaviour of the cars. Fast cars can pass the slow ones 
using the passing-lane. We write and solve the mean field rate equations for the 
density of slow and fast cars respectively. We also investigate the properties of 
the model through computer simulations and obtain the fundamental diagrams 
. A comparison between our results and $v_{max}=2$ version
of Na-Sch model is made.

\end{abstract}



\end{titlepage}

\newpage

\section{Introduction}

In recent years, modeling traffic flow has been the subject of
comprehensive studies by statistical physicists
\cite{schad2000,tgf96,tgf98,tgf2000,helbbook}. Needless to
say  many general phenomena in vehicular traffic can be explained in
general terms with these models. Distinct traffic states have been
identified and some of these models have found empirical applications in
real
traffic \cite{tgf96,tgf98,tgf2000}.
In these investigations,
various theoretical approaches namely microscopic car-following models
\cite{santen28,santen32}, hydro-dynamical coarse-grained
macroscopic
models \cite{helbbook,santen17,kim}, and gas-kinetic
models \cite{prigogin,helb-trib}
have been developed in order to find a better quantitative as
well as qualitative understanding toward vehicular traffic phenomena.
Recently as an alternative microscopic description, Probabilistic
Cellular Automata (PCA) have come into play (for an overview see Refs.
\cite{ito95,schad99EPJ} ). This approach to theoretical
 description of traffic flow
is one of the most effective and well-established ones and there is a
relatively
rich amount of results both numeric and analytic in the literature
\cite{schad2000,santen}.\\

 In PCA models, space (road), time and velocities of vehicles
are
assumed to take discrete values. This realization of traffic flow provides
PCA as an ideal tool for the computer simulation. One of the prototype
PCA
models
is the so-called {\it Nagel-Schreckenberg} (Na-Sch) model \cite{NS92}
which
describes a single-lane traffic flow. Although the initial observations of
the Na-Sch
model were numerical, shortly thereafter analytical technique were
also proposed 
\cite{ito95,schad99EPJ,santen}. Analytical treatments to CA are difficult
in
general.
This is mainly due to the discreteness and the use of parallel
(synchronous) updating procedures which produce the largest correlation
among the vehicles with regard to other updating schemes.\\
Soon after its introduction, the Na-Sch model was extended to account
for more realistic situations such as multi-lane traffic flow
\cite{rickert,nagel98},
bi-directional roads \cite{simon98} and urban traffic
\cite{esser,schad-cho}. In
multi-lane traffic,
fast cars are capable of passing the slow ones by using the fast-lane.
The
possibility of lane-changing allows for these models to exhibit
non-trivial and interesting properties which are exclusive to multi-lane
traffic flow. Despite the quite large approximative methods applied to
single-lane Na-Sch based models, there are few analytical approaches to
multi-lane traffic flow \cite{Nagatini}. One main reason is the large
number of rules in
PCA modeling multi-lane traffic. In reality, a driver attempting to
overtake the car ahead (in a uni-directional road) has to take the
following criteria into consideration:\\
1) There must be enough forward space in the passing-lane\\
2) There must be enough backward space in passing-lane so that no
accident could occur between two simultaneously passing cars.\\
 
Moreover, in bi-directional roads additional criteria are necessary for a
successful passing (for details see \cite{simon98} ).
The main purpose of the present paper is to introduce an analytical
approach to study a uni-directional two-lane road. The approach we use is
to some extent
similar to PCA, however basic differences are distinguishable. The major
distinction is concerned with the type of updating scheme. In contrast to
PCA which are realized in parallel update, our models are
based on time-continuous random sequential update. The mechanism of
modeling the two-lane traffic we use, is based on the stochastic
reaction-diffusion processes, however the rules have roots in the Na-Sch
rules.
This paper is organized as follows: In section two we define the first
model ( model I ) and interpret the rules in terms of those in Na-Sch
model. Section three starts with the Hamiltonian description of the related
master equation and continues with mean field rate equations and their
solutions. The results of the numerical simulation of the model I ends
this
section. Next we introduce the second model (model II) in section four 
which is formulated in symmetric as well as asymmetric versions
and
follow the same steps performed in section three to obtain the fundamental 
diagrams of the both versions. The paper ends with some concluding remarks
in section five.
\section{Definitions of the Models  }
In the first model, a uni-directional two-lane road is approximated by a 
set of two parallel one dimensional chains, each with $N$ sites. 
The periodic boundary condition applies to both. Cars are considered as 
particles which occupy sites of the chains. Two type of cars 
exist in the system: slow cars
 which are denoted by $A$ and fast cars denoted by $B$. Also $\Phi $ represents
 an empty site. Each site of the chains is either empty, occupied by a slow
 or by a fast car.\\
Fast cars can pass the slow ones with certain probabilities while 
approaching them. The bottom lane is the home-lane and cars are only allowed
to use the top lane for passing. Once the passing process is achieved, they 
should return to the home-lane. This realization of a two-lane road is 
regarded as "{\it asymmetric }" type. Nonetheless "{\it symmetric }" type
could also be implemented where passing from the right is allowed as
well.\\
In model I, we restrict ourselves to "{\it asymmetric }" type. The
state
of the system is characterized by two sets of occupation numbers 
$(\xi_1,\xi_2,\cdots,\xi_N)$ and $(\sigma_1,\sigma_2,\cdots,\sigma_N)$
for the home and passing-lane respectively. $\xi_i , \sigma _i  =0,1,2 $
where zero refers to an empty site whereas one and two refer to a site
being occupied by a slow or a fast car respectively.\\
To investigate the characteristics of this model, a simplification has been
considered. If simultaneous two-car occupation of parallel sites of the chains
is forbidden, one can describe configurations with a single set of occupation
numbers $\{ \tau_i \}$ where $ \tau_i =0,1,2 $.\\
Inspired by the $v_{max} =2 $ version of the Na-Sch model
\cite{NS92,ito95}, we
propose
the following set of stochastic processes which evolve according to a random 
sequential updating scheme:\\
\be
A\Phi \rightarrow \Phi A \ \ ( {\rm with} \ \ {\rm  rate }\ \  h )
\ee

\be
B\Phi \rightarrow \Phi B \ \ ( {\rm with} \ \ {\rm rate }\ \  p )
\ee

\be
A\Phi \rightarrow \Phi B \ \ ( {\rm with} \ \ {\rm rate }\ \  q )
\ee

\be
B\Phi \rightarrow \Phi A \ \ ( {\rm with} \ \ {\rm rate }\ \  r )
\ee

\be
B A \rightarrow  A A   \ \ ( {\rm with} \ \ {\rm rate }\ \  \lambda )
\ee

\be
B A \Phi \rightarrow \Phi A B    \ \ ( {\rm with} \ \ {\rm rate } \ \ s )
\ee

In order to illustrate the above definitions, let us express their 
interpretations :\\

The first and the second of the above rules correspond to the free moving
of
slow and fast cars respectively. The third one expresses the accelerated 
movement of slow cars. This step corresponds to the so-called acceleration 
step in the Na-Sch model. The fourth rule simulates the behaviour of a 
driver randomly 
reducing his/her speed as a result of environmental effects, road
conditions 
etc. This step corresponds to the so-called "{\it random breaking}" step
in Na-Sch model.
Finally the last two processes simulate the behaviour of the fast-car
drivers
when approaching a slow car. Either they pass the slow car using the 
passing-lane or they prefer to move behind it which give rises to their
speed reduction.\\

We recall that in Na-Sch model, the forward movement of each car is highly
affected by the car ahead. Here for simplicity we have considered the
two-site interactions and only use three-site interaction for the passing
process. In this particular case, it is crucial that the site ahead of the
slow car should be empty. 
Despite
the partial explanation of microscopic rules necessary for the description
of a traffic flow in a two-lane road, the present model ignores the effect of 
oncoming fast cars (in the passing-lane) on the fast car (in the
home-lane). In 
reality, a fast car attempts to overtake provided that there is enough back-
space behind him in the passing-lane i.e. there is no passing car close to
him
in the passing-lane \cite{rickert,nagel98}. In model (I) passing occurs
locally and irrespective
of the
state of passing-lane behind the fast car in the home-lane.\\
\section{Master Equation and Mean-Field Rate Equations}

The processes (1) to (6) could be regarded as a two-species one-dimensional
reaction-diffusion stochastic process. This is an example of hard-core
driven lattice gas far from equilibrium which has proven to be excellent
systems for theoretical investigations of low dimensional systems out of
thermal equilibrium. A large variety of phenomena had already been
described by driven lattice gases ( for an overview see \cite{schutzbook,
privbook} and the references therein). Using the rates given by (1-6), one can 
rewrite the corresponding master equation as a Schr\"odinger-like equation
in imaginary time.\\
\be
{\partial \over \partial t } \vert p(t) \rangle = -{\cal H} \vert p(t) 
\rangle
\ee

The explicit form of ${\cal H}$ could be written down via the rate
equations. Let $\langle n_{k,A} \rangle $ $( \langle n_{k,B} \rangle) $
denotes the
probability that at time $t$, the site $N=k$ of the chain is occupied by a
slow (fast) car. The Hamiltonian formulation of master equation allows for
evaluating the average
quantities in a well-established manner. It could be easily verified that
the following rate equations hold for the average occupation
probabilities.\\
\be
{d \over dt} \langle n_{k,A} \rangle = h \langle n_{k-1,A} e_k  
\rangle + r\langle n_{k-1,B} e_k \rangle   
 + \lambda \langle n_{k,B}
 n_{k+1,A} \rangle  -h\langle n_{k,A} e_{k+1} \rangle
-q\langle n_{k,A} e_{k+1} \rangle 
\ee

In the above equation $e_k$ stands for $1-n_{k,A} -n_{k,B} $. Similarly
for $\langle n_{k,B} \rangle $ we have:\\
\be
{d \over dt} \langle n_{k,B} \rangle = q \langle n_{k-1,A} e_k  
\rangle + p\langle n_{k-1,B} e_k \rangle + s\langle 
n_{k-2,B} n_{k-1,A}e_k \rangle  -r\langle n_{k,B} e_{k+1} \rangle
-p\langle n_{k,B} e_{k+1} \rangle  
\ee
$$
 - \lambda \langle n_{k,B}
 n_{k+1,A} \rangle - s\langle n_{k,B} n_{k+1,A} e_{k+2} \rangle
$$\\
Apparently the total number of neither slow nor fast cars are conserved 
according to the dynamics and therefore the right hand side of eqs (8,9)
cannot be written as a difference of two currents. However, the total number
of cars i.e. the sum of slow and fast cars is a conserved quantity and
the time rate of changing $ \langle n_{A,k} \rangle  + \langle n_{B,k} 
\rangle $ is equal to a difference between oncoming and outgoing currents.
Summing up
eqs (8) and (9) yields the following discrete form of the continuity
equation:\\
\be
{d \over dt} \left[ \langle n_{k,A} \rangle + \langle n_{k,B} 
\rangle \right] = \langle J_{k}^{in} \rangle  - \langle 
J_{k}^{out} \rangle
\ee

In which the explicit form of $ \langle J_{k}^{out} \rangle $ is given
below. \\
\be
\langle J_{k}^{out} \rangle = h\langle n_{k,A} e_{k+1} \rangle
+r\langle n_{k,B} e_{k+1} \rangle + q\langle n_{k,A} e_{k+1} 
\rangle +p\langle n_{k,B} e_{k+1} \rangle + s\langle n_{k,B} n_{k+1,A}
e_{k+2}
\rangle
\ee \\
 Equations (8), (9) and (11) are
valid for arbitrary time $t$, however our particular interest is focused in the
longtime behaviour of the system where stationarity is established. In the 
steady state regime, one and two-points correlators in (8,9) will be
time-independent. Equation (10) implies that in steady state the current
would
be site-independent as expected.\\ 
So far, our result have been exact and no approximation has been implemented. 
At this stage and in order to solve equation (8-11) we resort to a
mean-field
approximation and replace the two point correlators with the product of 
one-point correlators. Moreover, since the closed boundary condition has
been
been applied, it can be anticipated that the steady values of 
$ \langle n_{k,A} \rangle _s $ and $\langle n_{k,B} \rangle _s$
be site-independent and therefore we omit the site-dependence subscripts
from equations (8-11). Denoting the steady values of
$ \langle n_A \rangle _s $ and $\langle n_B \rangle _s$ by
$n_A$ and $n_B$ respectively, the steady  current $J$ turns out to be \\  
\be
J=(hn_A +rn_B +qn_A + pn_B + sn_A n_B)(1-n)
\ee

In the above expression, the total density of the cars has been taken 
to be $n$ \\
\be
n_A +n_B =n
\ee
our final aim is to write $J$ in terms of total density $n$ and the rates 
. This is performed if one writes $n_A$ as a function $n$ 
and the rates. By applying the mean-field approximation to the equation
(9)
in its steady state form and using (13), one obtains the following
equation\\
\be
r(n-n_A)(1-n) + \lambda (n-n_A)n_A = qn_A(1-n)
\ee
which simply yields the solutions:\\
\be
n_A ={1 \over {2\lambda } } \left[ n\lambda -(1-n)(q+r) \pm \left( [
n\lambda 
-(1-n)(q+r)]^2 + 4rn(1-n)\lambda \right)^{ {1 \over 2 } } \right]
\ee
the solution with the minus sign is unphysical $(n_A <0 )$ so the unique 
solution is the one with the positive sign.
We Remark that within the mean-filed approach, one also can solve the
time-dependent version of the equations (8,9). In this case, the equation
for $\langle n_A \rangle $ turns out to be:\\
\be
{d  \over dt } \langle n_A \rangle = rn(1-n)- [ (q+r)(1-n)-n\lambda
] \langle n_A
\rangle -\lambda (\langle n_A \rangle)^2
\ee
 which simply give rises to the following solution:\\
\be
\langle n_A \rangle (t) =  { n_A -C_1 e^{ C_2 (C_3 -t) } \over 1- e^{C_2
(C_3 -t ) } }
\ee

In which $ C_1, C_2 $ and $C_3$ are constants depending on the rates. In
the long-time limit, the mean concentration of slow cars exponentially
relaxes toward the steady value $n_A$.
 Replacing the above $n_A$ into the 
equation (13), one now has the total current $J$ as a function of $n$ and
the rates. In order to have better insights into the problem, extended
computer simulation
were carried out. Here we present the result of numerical investigations
of
model I. In these computer simulations, the system size is typically 2400.
With no loss of generality, we re-scale the time so that the rate of
hopping a fast car is set to one.
 The speed of slow cars is supposed to be 70 percent of the speed of 
 the fast cars which is realized by taking $h=0.7$ . The values of
$q$ and $\lambda $ are set 1 and 0.7 respectively.
 One {\it sub-update } step  consists of a random selection of a site,
say $N=i$ and developing the state of the link $(i,i+1)$ according to the 
 dynamics. One {\it update } step contains $L$ {\it sub-updates }. The typical
 number of updates developed in order that the system reaches to
stationarity is 400000 and the averaging has been performed over 500000
updating steps.
 The initial state of the system was prepared randomly i.e. each site is
occupied with the probability $n$. figures below show the result of
numerical simulations.\\

\begin{figure}
\centerline{\psfig{figure=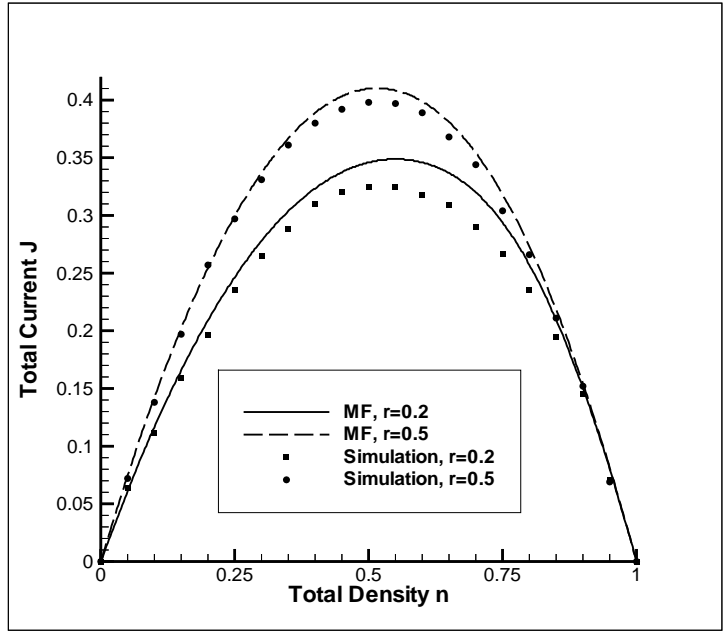,width=.6\columnwidth}}
\vspace{5 mm }
\caption{ current-density diagram for different values of r. s is set to 
          0.4 .               }
\end{figure}

\begin{figure}
\centerline{\psfig{figure=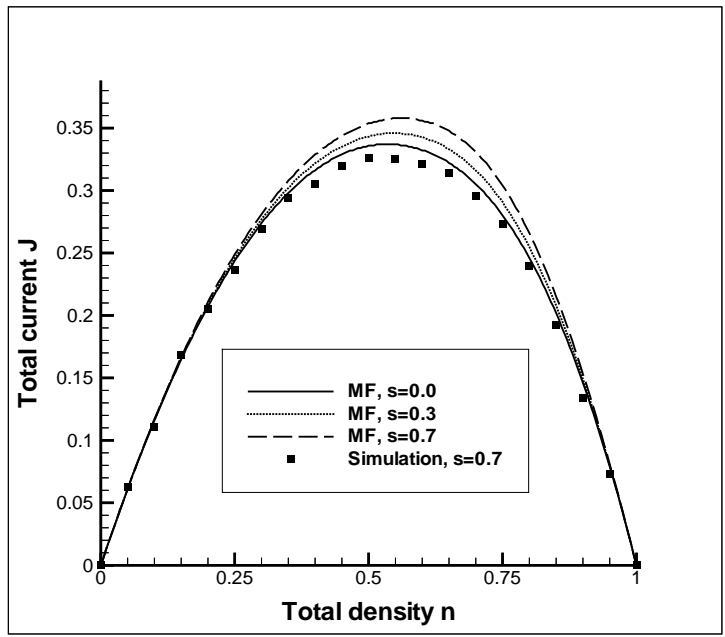,width=.6\columnwidth}}
\vspace{5 mm }
\caption{ current-density diagrams for different values of s. r is set to
0.2 . }
\end{figure}

\begin{figure}
\centerline{\psfig{figure=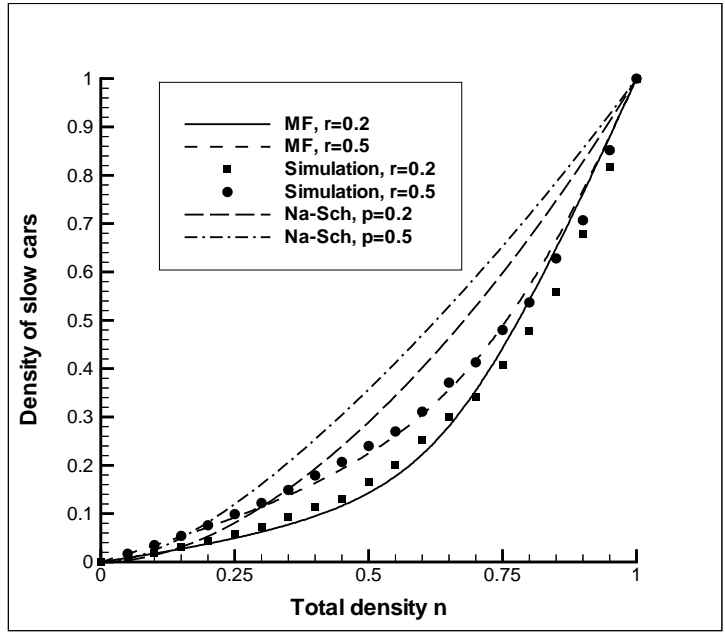,width=.6\columnwidth}}
\vspace{5 mm }
\caption{  density of slow cars versus the total density for s
= 0.4 . }
\end{figure}

\begin{figure}
\centerline{\psfig{figure=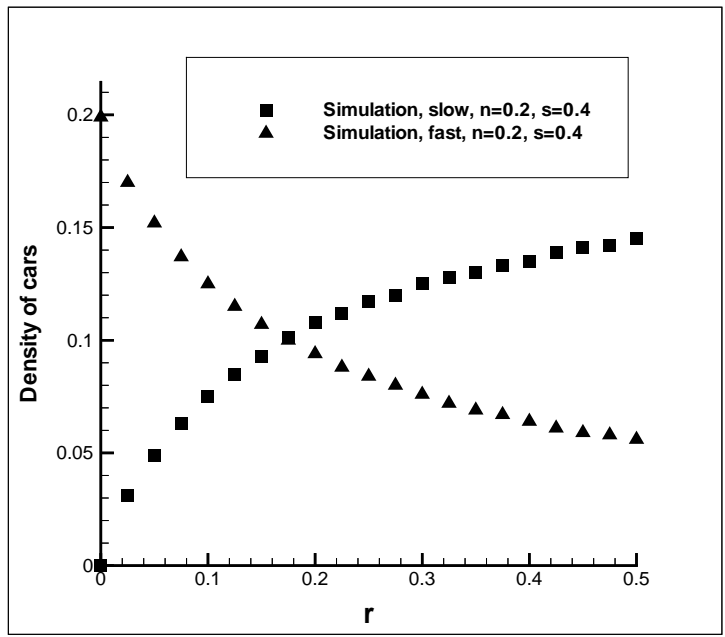,width=.6\columnwidth}}
\vspace{5 mm }
\caption{ density of slow (fast) cars as a function of r . The values of
$n$ and $s$ are 0.2 and 0.4 respectively. }
\end{figure}



\section{Model II}
\subsection{Asymmetric regulation}

The second model we consider, has less resemblance to the Na-Sch model.  
Here there is no specification of fast and slow cars and only one kind of 
particle exists in the chain, nevertheless the distinction between fast and 
slow cars is realized by their appearance in the passing and home-lanes.
In
this periodic double-chain model the following processes occur in a
random sequential updating scheme :\\

$ \circ ~~ \circ ~~~~~~~~~~~~~~~~~~~~~~~~~~~~~~~~~~~~~~~ \circ ~~ \circ
$\\
$
~~~~~~~~~ \longrightarrow  ~~~~~~~ {\rm with } \ \ {\rm rate }  \ \ h   
$\\
$ ~~~~~\bullet ~ \circ~~~~~~~~~~~~~~~~~~~~~~~~~~~~~~~~~~~~~~~ \circ ~~
\bullet $\\

$ \circ ~~ \circ ~~~~~~~~~~~~~~~~~~~~~~~~~~~~~~~~~~~~~~~ \circ ~~ \bullet
$\\
$
~~~~~~~~~ \longrightarrow  ~~~~~~~ {\rm with } \ \ {\rm rate }  \ \ a   
$\\
$ ~~~~~\bullet ~ \bullet~~~~~~~~~~~~~~~~~~~~~~~~~~~~~~~~~~~~~~~ \circ ~~
\bullet $\\

$ \bullet ~~ \circ ~~~~~~~~~~~~~~~~~~~~~~~~~~~~~~~~~~~~~~~ \circ ~~ \circ
$\\
$
~~~~~~~~~~ \longrightarrow  ~~~~~~~ {\rm with } \ \ {\rm rate }  \ \ g   
$\\
$ ~~~~~\bullet ~ \circ ~~~~~~~~~~~~~~~~~~~~~~~~~~~~~~~~~~~~~~~ \bullet ~~
\bullet $\\

$ \bullet ~~ \circ ~~~~~~~~~~~~~~~~~~~~~~~~~~~~~~~~~~~~~~~ \circ ~~
\bullet
$\\
$
~~~~~~~~~~~ \longrightarrow  ~~~~~~~ {\rm with } \ \ {\rm rate }  \ \ b   
$ \\
$ ~~~~~\bullet ~ \bullet~~~~~~~~~~~~~~~~~~~~~~~~~~~~~~~~~~~~~~~ \bullet
~~
\bullet $\\

As depicted, the "{\it asymmetric }" regulation has been adopted so that
the top-lane can only be used for passing. According to the above rules,
once a successful passing 
has taken place, the passing car should return to its home-lane unless the 
next site in the home-lane is already occupied. In this circumstance, it
can
continue to pass the second slow car ( multi-passing ). Each site of the 
double chain takes four different states but according to the above
dynamics
only three of them appears in the course of time. The forbidden state is
the one
in which the passing-lane site is full and its parallel home-lane site is
empty. 
 Regarding this fact, we characterize the three allowed states by $\Phi ,
 A $ and $B$. $\Phi$ represents the situation where both parallel sites
are empty, $A$ represents the case of an occupied site in the home-lane
and empty
parallel site in passing-lane and finally $B$ refers to the case of both 
parallel sites being occupied.

This notation yields the following reaction-diffusion processes:\\

\be
A~ \Phi \rightarrow \Phi ~A  \ \ \ ({\rm with } \ \ {\rm rate} \ \ h ) 
\ee                                
\be
A~A \rightarrow \Phi ~B  \ \ \ ({\rm with } \ \ {\rm rate} \ \ a ) 
\ee                                
\be
B~ \Phi \rightarrow A~A  \ \ \ ({\rm with } \ \ {\rm rate} \ \ g ) 
\ee                                
\be
B~A \rightarrow  A~B  \ \ \ ({\rm with } \ \ {\rm rate} \ \ b ) 
\ee                                

It is worth mentioning that the above model for a two-lane road is
simultaneously being considered within the approach of Deterministic 
Cellular Automata (DCA) \cite{krug}.

\subsection{Master equation and mean-field approach}

Similar to the steps performed in model I, one can write the following
 form of discrete continuity equation.\\
\be
{d \over dt} \left[ \langle n_{k,A} \rangle + 2\langle n_{k,B} \rangle
\right] =\langle J_{k}^{ in } \rangle - \langle J_{k}^{ out
} \rangle
\ee
in which\\
\be
\langle J_{k}^{ out} \rangle = h\langle n_{k,A} e_{k+1} \rangle 
+b \langle n_{k,B} n_{k+1,A} \rangle + g\langle n_{k,B} e_{k+1} \rangle
+ a\langle n_{k,A} n_{k+1,A} \rangle
\ee \\
The above expression for $\langle J_k \rangle $ has a clear interpretation 
in terms of rules (18-21).
In steady state, the time dependences in the equation disappear
and the current will be site-independent.
Next we apply the mean-field approximation through which all the two-point
correlators are replaced by the product of one-point correlators. This
leads to the
following equation for $J$:\\
\be
J=hn_A(1-n) +b ( n-{n_A \over 2 })n_A + g(n-{n_A \over 2})(1-n) +
an_A^2
\ee
where the relation $ {n_A \over 2 }+n_B = n $ has been used.\\
In order to obtain $J$ in terms of total density $n$ and the rates, we must
write $n_A$  as a function of $n$ and the rates. This is done by solving the
following equation with its left hand side set to zero.\\
\be
{d \over dt } n_A = 2gn_B (1-n) -2an_A^2 
\ee
The unique physical solution of the above equation is:\\
\be
n_A = {1 \over 4a } \left[ \{ ( g^2(1-n)^2 +16an(1-n)g ) \}^{{1 \over 2 }}
 -g(1-n) \right]
\ee

putting (26) in the eq. (24),  the current $J$ is now obtained in terms of
$n$ and
the rates. The result of computer simulations are shown in the following
set of figures.
Here the rate $b,g$ and $h$ are chosen to be $1.0 ,1.0 $ and $0.7$ 
respectively while $a$ is varied. We recall that "$a$" measures the
tendency
of fast cars to pass the slow ones.The simulation specifications are the
same as those in model I.\\

\begin{figure}
\centerline{\psfig{figure=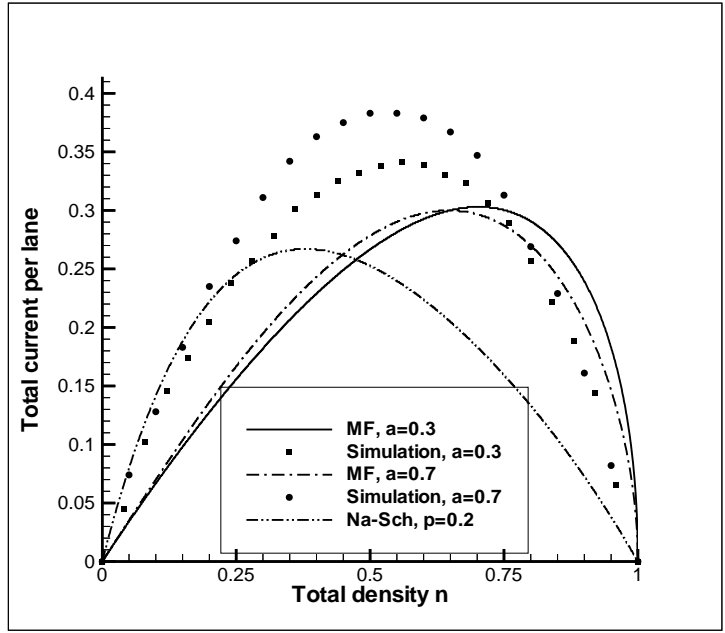,width=.6\columnwidth}}
\vspace{5 mm }
\caption{ current-density diagram for different values of passing rate. 
}
\end{figure}

\begin{figure}
\centerline{\psfig{figure=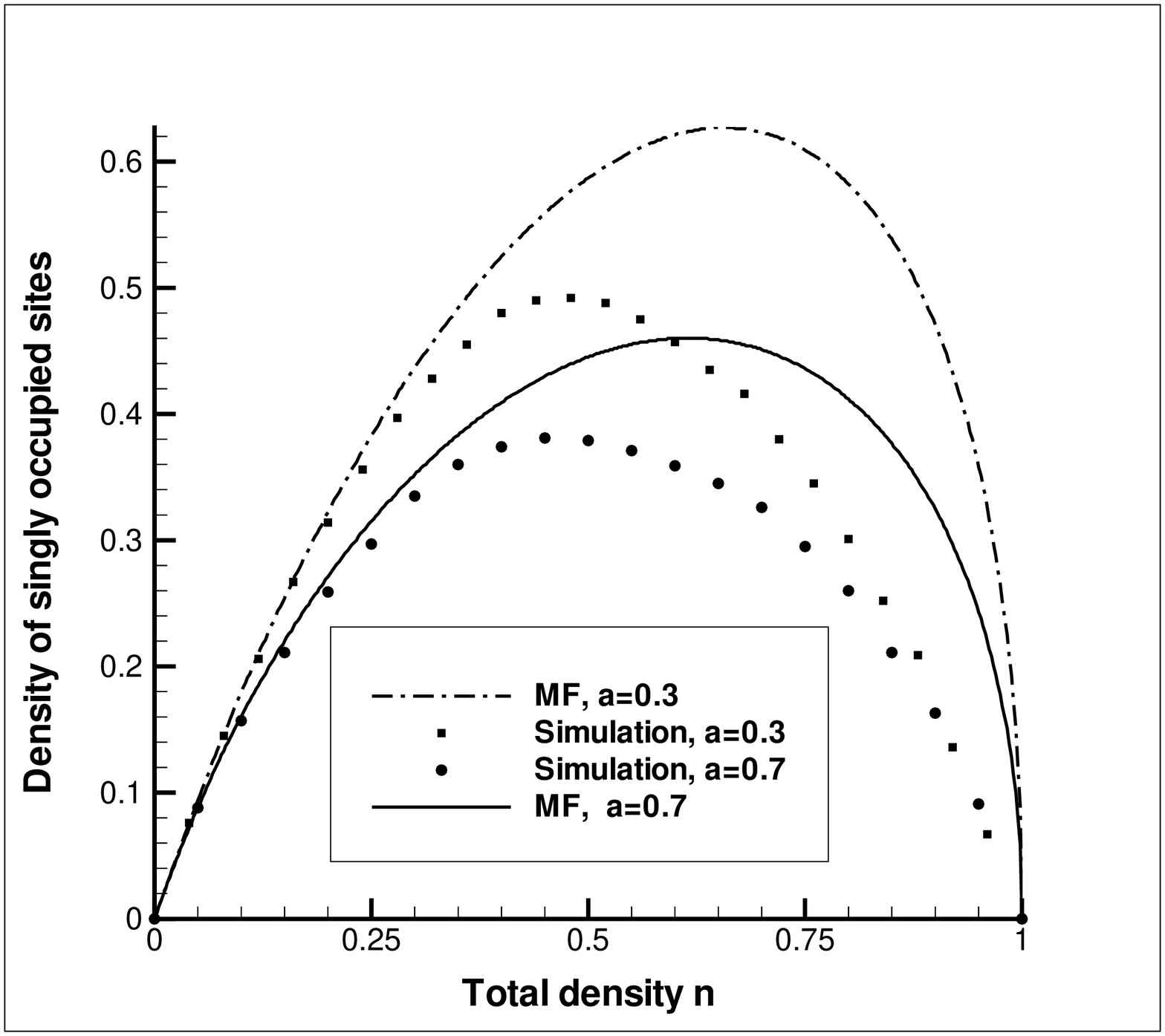,width=.6\columnwidth}}
\vspace{5 mm }
\caption{ density of singly occupied sites versus the total density 
}
\end{figure}

\begin{figure}
\centerline{\psfig{figure=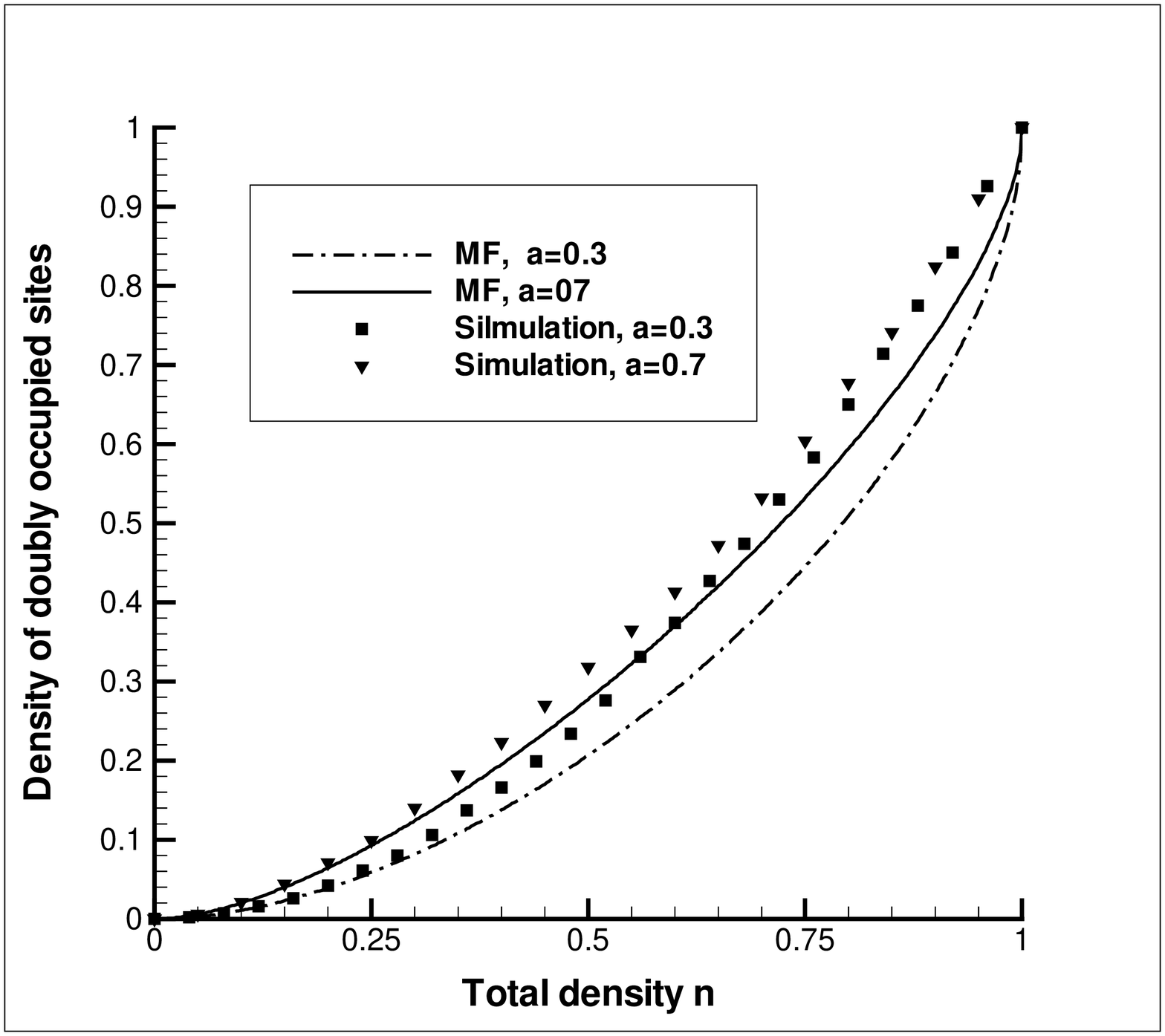,width=.5\columnwidth}}
\vspace{5 mm }
\caption{ density of doubly-occupied sites versus the total density. }
\end{figure}

\subsection{Symmetric Regulation}

Here we allow the fast cars to pass rightward as well. In this case, both
the top and bottom lanes become identical and fast cars can pass the slow
ones irrespective of their home-lane. In this {\it symmetric } two-lane
model, each particle hops one site ahead in its home-lane provided that
the next site is empty. Otherwise it tries to pass the car ahead. This 
attempt is successful if there is an empty site ahead on the opposite
lane. The following rules illustrates the model definition.\\
 
$ \ast ~~ \ast ~~~~~~~~~~~~~~~~~~~~~~~~~~~~~~~~~~~~~~~ \ast ~~ \ast
$\\
$
~~~~~~~~~ \longrightarrow  ~~~~~~~ {\rm with } \ \ {\rm rate }  \ \ h   
$\\
$ ~~~~~\bullet ~ \circ~~~~~~~~~~~~~~~~~~~~~~~~~~~~~~~~~~~~~~~ \circ ~~
\bullet $\\

$ \bullet ~~ \circ ~~~~~~~~~~~~~~~~~~~~~~~~~~~~~~~~~~~~~~~ \circ ~~
\bullet
$\\
$
~~~~~~~~~ \longrightarrow  ~~~~~~~ {\rm with } \ \ {\rm rate }  \ \ h   
$\\
$ ~~~~~\ast ~ \ast~~~~~~~~~~~~~~~~~~~~~~~~~~~~~~~~~~~~~~~ \ast ~~
\ast $\\

$ \bullet ~~ \bullet ~~~~~~~~~~~~~~~~~~~~~~~~~~~~~~~~~~~~~~~ \circ ~~
\bullet
$\\
$
~~~~~~~~~~ \longrightarrow  ~~~~~~~ {\rm with } \ \ {\rm rate }  \ \ g   
$\\
$ ~~~~~\circ ~ \circ ~~~~~~~~~~~~~~~~~~~~~~~~~~~~~~~~~~~~~~~ \circ ~~
\bullet $\\

$ \circ ~~ \circ ~~~~~~~~~~~~~~~~~~~~~~~~~~~~~~~~~~~~~~~ \circ ~~
\bullet
$\\
$
~~~~~~~~~~~ \longrightarrow  ~~~~~~~ {\rm with } \ \ {\rm rate }  \ \ g
$ \\
$ ~~~~~\bullet ~ \bullet~~~~~~~~~~~~~~~~~~~~~~~~~~~~~~~~~~~~~~~ \circ
~~
\bullet $\\
The {\it astrix } symbol indicates that the process in the opposite lane
occurs independently of the configuration of the sites filled with {\it
astrix}. If we denote the state of two parallel sites in which the bottom
site is empty and the top one is occupied by $B$, the state of
 simultaneous occupation of parallel
sites by $C$
and adopting the notations $\Phi$ and A as the same in the
asymmetric version of the model, then it could easily be verified that the
forms of the discrete continuity equation and the current are as
follows:\\
\be
{d \over dt} \left( \langle a_{k} \rangle + \langle b_{k} \rangle
+2\langle c_{k} \rangle \right) =\langle J_{k-1}\rangle - \langle J_{k}
\rangle
\ee
and\\
$
\langle J_{k,k+1} \rangle = h \left( \langle a_{k} e_{k+1} \rangle 
+ \langle a_{k} b_{k+1} \rangle + 2\langle c_{k} e_{k+1} \rangle
+ \langle c_{k} b_{k+1} \rangle
+ \langle b_{k} e_{k+1} \rangle + \langle b_{k} a_{k+1} \rangle
+ \langle c_{k} a_{k+1} \rangle \right)
$\\
\be
 + g \left( \langle b_k b_{k+1}
\rangle + \langle a_k a_{k+1} \rangle \right)
\ee

Where $\langle a_k \rangle , \langle b_k \rangle $ and $ \langle c_k
\rangle $ refer to the probabilities that at time $t$, the site $N=k$
of the double-chain has one car in bottom lane, one car in top lane and
double-occupancy in both lanes respectively. In steady state, the system
is both time and site independent. Denoting the steady values of 
 $\langle a_k \rangle , \langle b_k \rangle $ and $ \langle c_k
\rangle $ by $a, b$ and $c$, one has the relation:
\be
{a+b \over 2 } + c = n
\ee
Moreover, the symmetry between the lanes implies that $a=b$.
The steady value $a$ is easily found to be obtained from the following
equation:\\
\be
(g+h)a^2 = hc(1-n)
\ee

Solving the steady-state equation for $a$, one finds:\\
\be
a= { ( \left[h^2(1-n)^2 +4hn(1-n)(g+h) \right] )^{{1 \over 2 }} -h(1-n)
\over 2(g+h) }
\ee

Also equation (31) leads to the following equation for $J$.
\be
J = 2 \left[ hn(1-n) + h \{ a^2 +2a(n-a) \}  + ga^2 )\right]
\ee 
Where by putting the eq. (31) into it, one reaches to expression for $J$
in terms of $n,g$ and $h$. We remark that the factor two reflects the
number of lanes. The result of computer simulations are shown in the
following set of figures. The value of $h$ is set to one and $g$ is
varied. 

\begin{figure}
\centerline{\psfig{figure=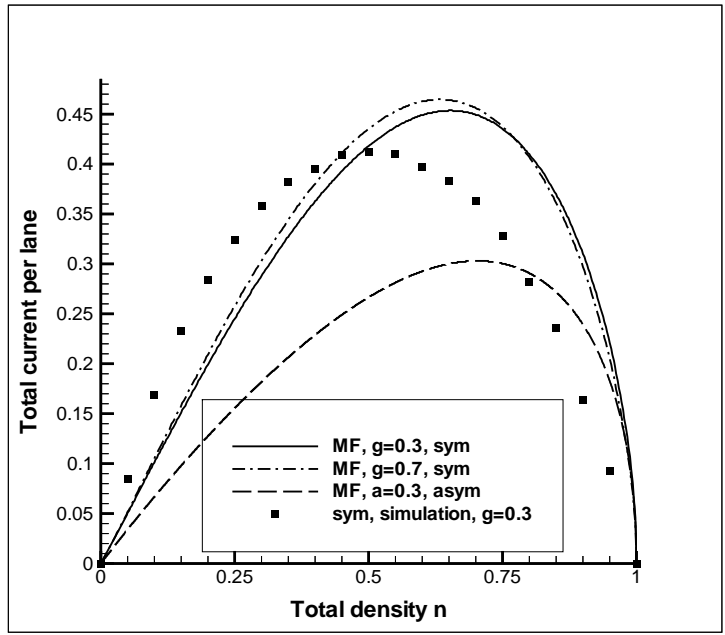,width=.6\columnwidth}}
\vspace{5 mm }
\caption{ current per lane-density diagram for different values of passing
rates }
\end{figure}

\begin{figure}
\centerline{\psfig{figure=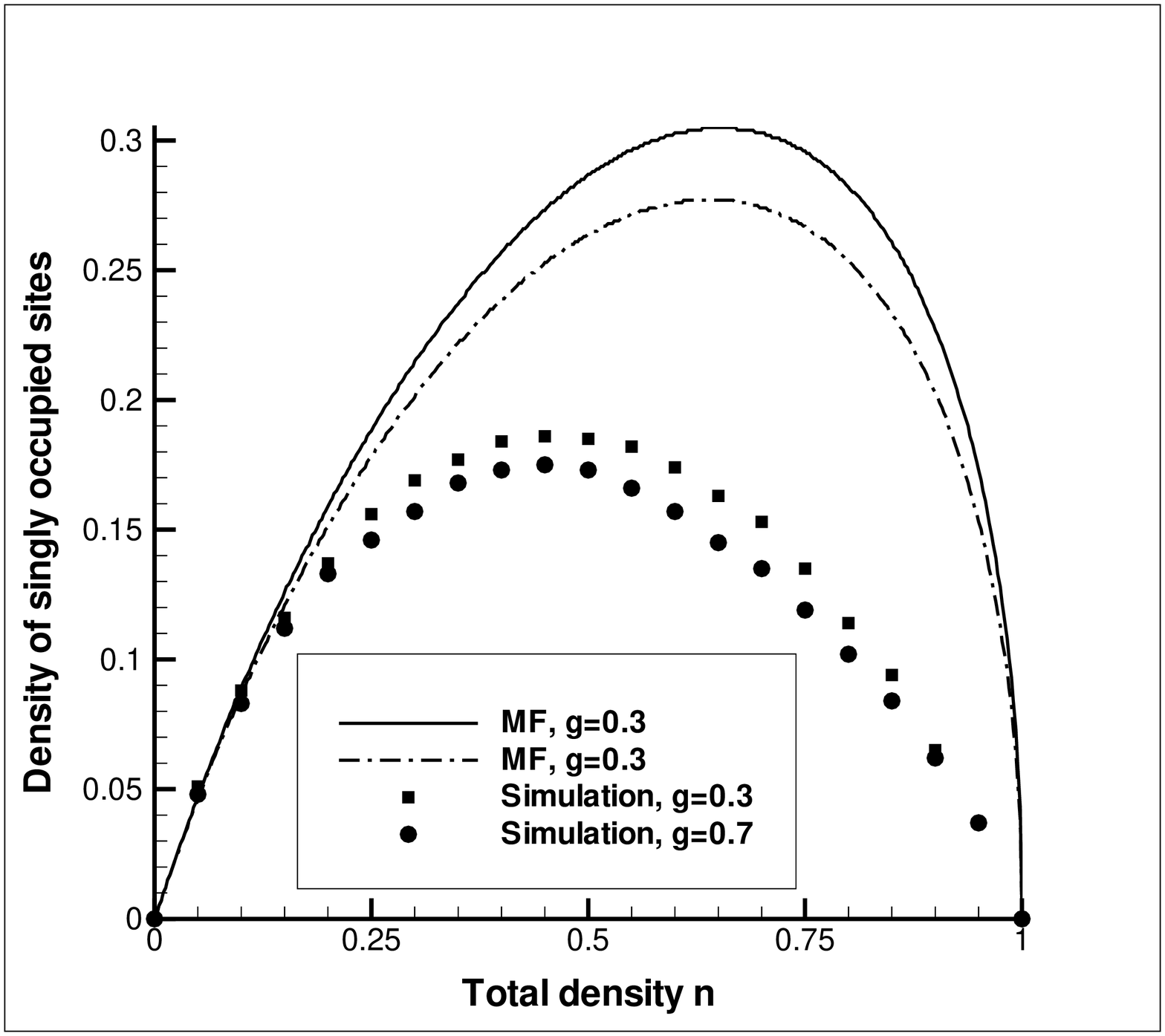,width=.6\columnwidth}}
\vspace{5 mm }
\caption{ density of singly occupied sites versus the total density. }
\end{figure}

\begin{figure}
\centerline{\psfig{figure=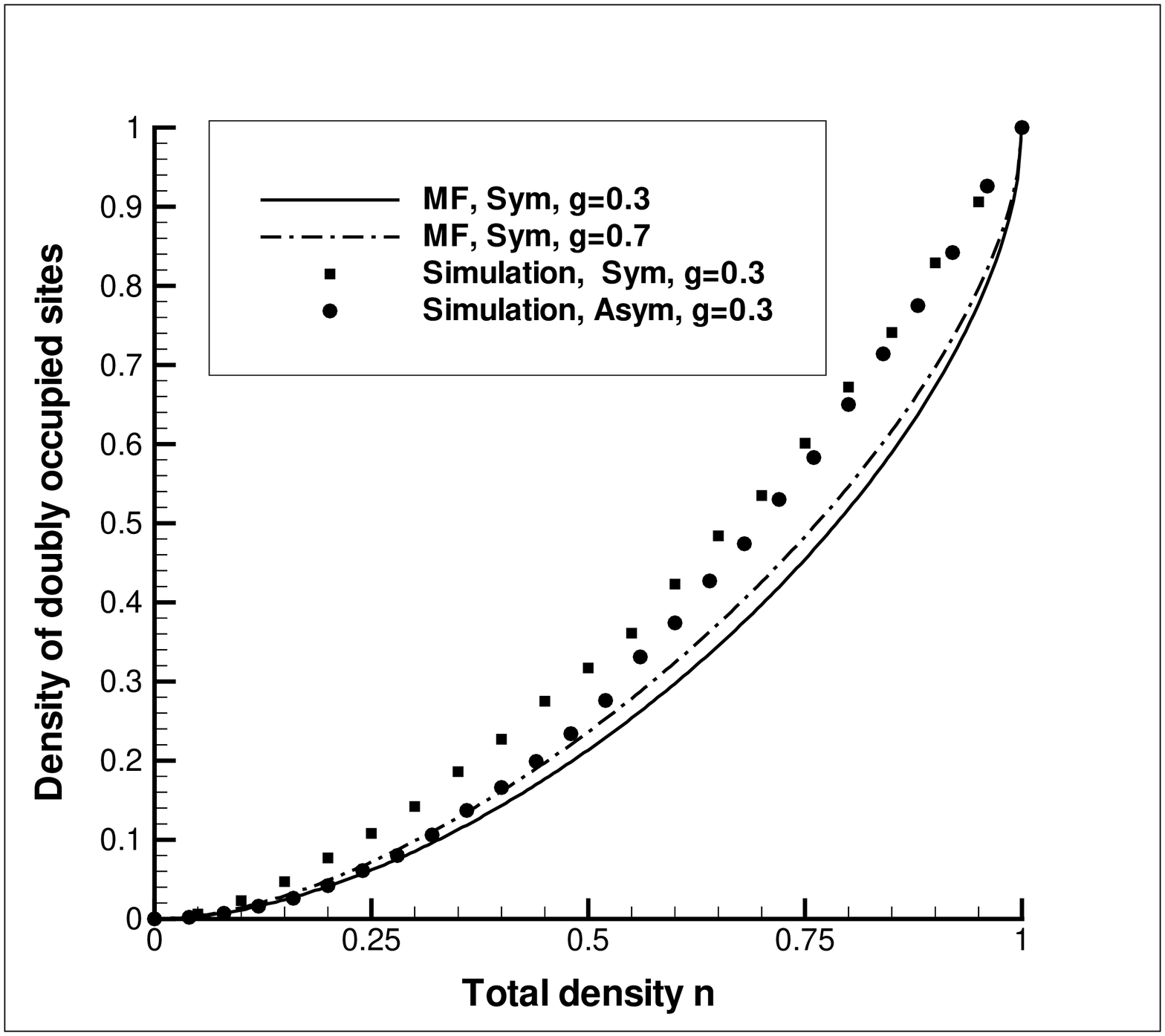,width=.6\columnwidth}}
\vspace{5 mm }
\caption{ density of doubly occupied sites versus the total density. }
\end{figure}

\begin{figure}
\centerline{\psfig{figure=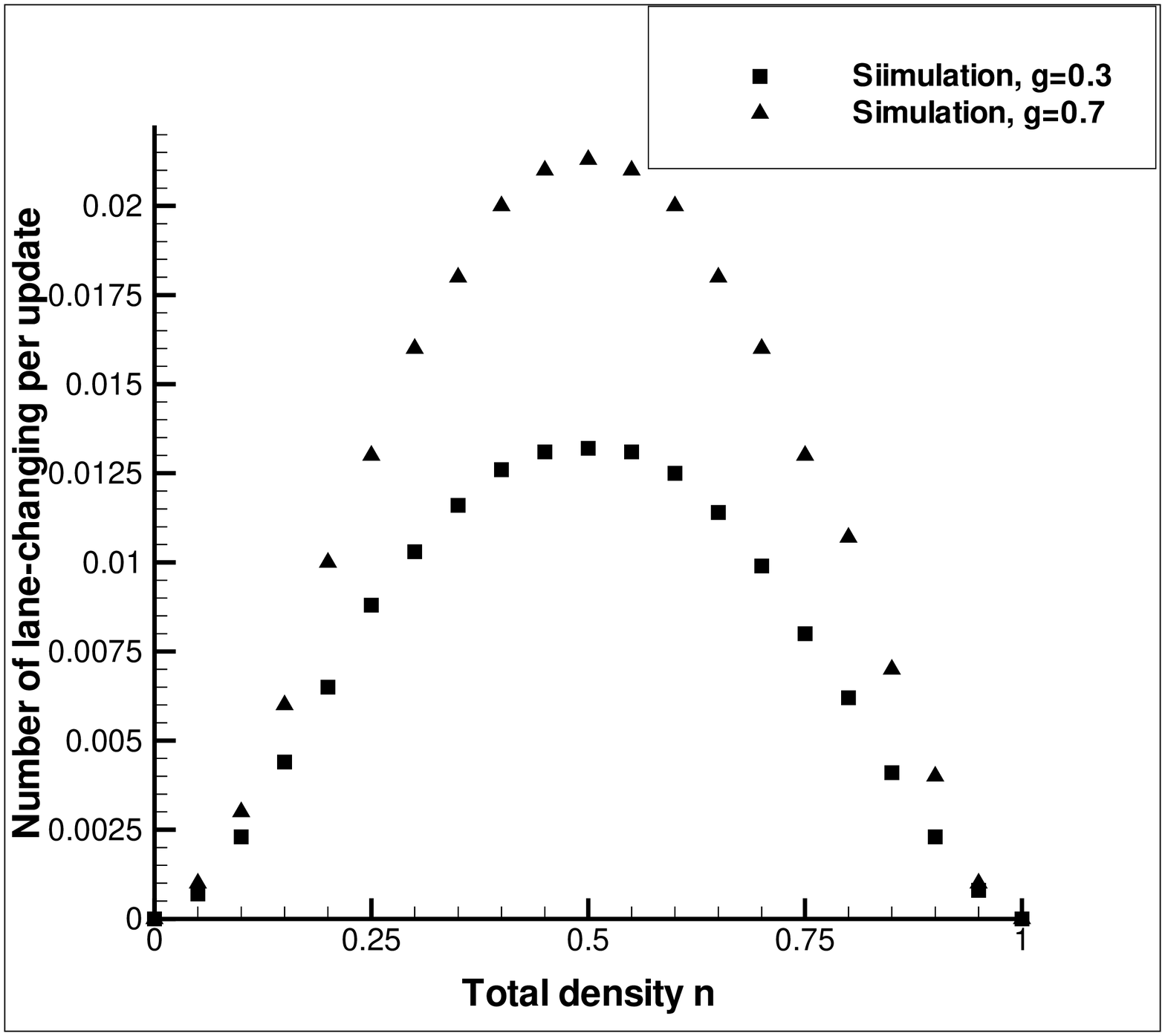,width=.6\columnwidth}}
\vspace{5 mm }
\caption{ number of lane-changing per update versus total density. }
\end{figure}

\section{ Concluding Remarks }
We have introduced a two-species reaction-diffusion model for description
of a uni-directional two-lane road. The type of update we have used is
random-sequential which sounds more appropriate for analytical treatments.
In the first model, the result of numeric simulations are very close to
those in mean-field approach which indicates that the effects of
correlations are suppressed. However in the second model, there are
remarkable differences between analytical and numeric results. in model I,
the current-density diagram is slightly affected by changing the passing
rate and the passing process has its most effect in the intermediate
densities. This could be anticipated since in the low and high densities
the number of passing considerably reduces. The space-time diagrams of the
model I reveal the discriminating effect of passing.\\
In model II (both symmetric and asymmetric), the maximum of $J$ occurs in
different values of $n$ in simulation and analytical approach. Mean-field
predicts a shift toward higher densities while in simulation a slight
shift toward left is observed. We note that in the PCA based models the
maximum of $J$ corresponds to a considerable left-shifted value of the
density \cite{rickert,nagel98}. In symmetric version of the model II, we
observe an increment of the current with regard to the asymmetric
version. In contrast to the asymmetric version, the maximum of $J$ in
mean-field approach is higher than its value obtained through simulation.
Although the current diagram (10) appears asymmetrically with respect to
the density, the lane-changing diagram (13) is symmetric to a high
accuracy.\\

{ \large \bf Acknowledgements:} \\
I would like to express my gratitude to V. Karimipour and G. Sch\"utz,
 for their fruitful comments. 
I acknowledge the assistance given by A. Schadschneider, N.Hamadani,
V.Shahrezaei, R. Sorfleet and
in particular I highly 
appreciate R. Gerami for his valuable helps in the computer simulations.


\end{document}